\documentclass[showpacs,preprintnumbers,amsmath,amssymb]{revtex4}
\usepackage{graphicx}
\usepackage{dcolumn}
\usepackage{bm}
\begin{document}

\title{Phonon effect on two coupled quantum dots at finite temperature}
\author{Cheng-Ran Du}
\author{Ka-Di Zhu}

\affiliation{Department of Physics, Shanghai Jiao Tong University,
Shanghai 200240, People's Republic of China}

\begin{abstract}
The quantum oscillations of population in an asymmetric double
quantum dots system coupled to a phonon bath are investigated
theoretically. It is shown how the environmental temperature has
effect on the system.
\end{abstract}

\maketitle

 The asymmetric double semiconductor quantum dots system
(DQD) is referred as quantum dot molecule due to its similar
properties to natural molecule. By using self-assembled dot growth
technology\cite{self-assenbled} we can fabricate these molecule
-like dots, in which the confined electrons can transfer between
each quantum dot through tunneling effect. Up to now, such system
has been studied extensively. Under the influence of an external
oscillatory (optical pulse or voltage) driving field, one electron
can be excited from the valence to the conduction band in one dot,
which can in turn tunnel to the second
dot\cite{brazil}\cite{twodot}. Recent progress in semiconductor
nanotechnology indicate that these features have advantages for the
application in many quantum devices, such as QD
lasers\cite{laser}\cite{laserr}, QD diodes\cite{diodes} as well as
quantum computing processes\cite{comp}\cite{compp}. However, QDs are
embedded in the surrounding solid matrix, thus the effect of
electron-phonon interaction during the tunneling is not
negligible\cite{zhu}. Therefore, the environmental temperature will
have significant influence on the desired QD devices. In this
letter, we analyze how the optically driven asymmetrical DQD device
depends on the environmental temperature.
\begin{figure}
\begin{center}
\includegraphics[scale=0.4]{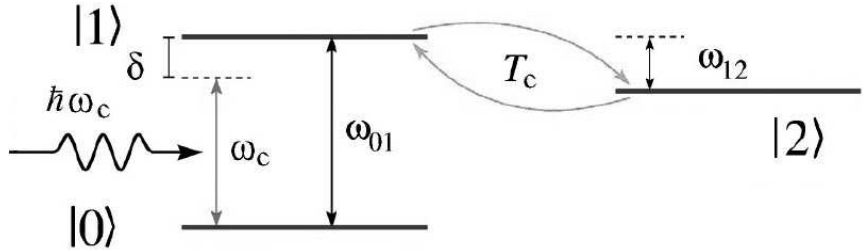}
\caption{Level configuration of a double QD system. A pulsed laser
excites one electron from the valence band that can tunnel to other
dot with the application of the voltage. We assume that the hole
cannot tunnel in the time scale we are considering
here.}\label{figure1}
\end{center}
\end{figure}

The asymmetrical DQD consists of two dots (the left and the right
one) with different geometries and has the ground state $|0\rangle$
( the system without excitation), first excited state $|1\rangle$ (
a pair of electron and hole bound in the left dot) and second
excited state $|2\rangle$ (one hole in the left dot and one electron
in the right dot). Any other states are effectively decoupled from
these three states because the valence band levels of two dots
become far off-resonance and the hole cannot tunnel to the right dot
consequently. Such model can be shown in Fig.\ref{figure}.
Considering that the single electron is unavoidably scattered by
phonons while tunneling between two dots, the Hamiltonian is given
by ($\hbar=1$)\cite{brazil}\cite{zhu}
\begin{equation}
\begin{split}
H=&\sum_{j=0}^2\varepsilon_{j}|j\rangle\langle{}j|+\Omega(e^{i\omega_{c}t}|0\rangle\langle1|
   +e^{-i\omega_{c}t}|1\rangle\langle0|)+T_{c}(|1\rangle\langle2|+|2\rangle\langle1|)\\
  &+\sum_{k=0}\omega_{k}b_{k}^{+}b_{k}
   +\frac{1}{2}(|1\rangle\langle1|-|2\rangle\langle2|)\sum_{k}g_{k}(b_{k}^{+}+b_{k}),
\end{split}
\end{equation}
where $\varepsilon_{j}$ is the energy of state$|j\rangle$, $T_{c}$
is the electron-tunneling matrix element, $\omega_{c}$ is the
frequency of the applied field,
$\Omega(t)=\langle0|\mu{}E(t)|1\rangle$, where $\mu$is the electric
dipole moment, describes the coupling to the radiation field of the
excitonic transition, $E(t)$ is the optical pulse amplitude.
$b_{k}^{+}(b_{k})$ and $\omega_{k}$ are the creation (annihilation)
operator and energy for $k$th phonon mode, respectively, $g_{k}$is
the coupling constant determined by the crystal material.

Applying a canonical transformation with the
generator\cite{zhu}\cite{zhuu}\cite{book}
\begin{equation}
S=(|1\rangle\langle1|-|2\rangle\langle2|)\sum_{k}\frac{g_{k}}{2\omega_{k}}(b_{k}^{+}-b_{k}),
\end{equation}
we have
\begin{equation}
H'=e^{S}He^{-S}=H'_{0}+H'_{I},
\end{equation}
where
\begin{equation}
H'_{0}=\varepsilon_{0}|0\rangle\langle0|+(\varepsilon_{1}-\Delta)|1\rangle\langle1|
+(\varepsilon_{2}-\Delta)|2\rangle\langle2|+\sum_{k}\omega_{k}b_{k}^{+}b_{k},
\end{equation}
\begin{equation}
H'_{I}=\Omega(e^{i\omega_{c}t}|0\rangle\langle1|e^{-A}+e^{-i\omega_{c}t}|1\rangle\langle0|e^{A}
+T_{c}(|1\rangle\langle2|e^{2A}+|2\rangle\langle1|e^{-2A}),
\end{equation}
where
\begin{equation}
\Delta=\sum_{k}\frac{g_{k}^{2}}{4\omega_{k}},
A=\sum_{k}\frac{g_{k}}{2\omega_{k}}(b_{k}^{+}-b_{k}).
\end{equation}

The Hamiltonian in the interaction picture is given by
\begin{equation}
\begin{split}
H''=&e^{iH'_{0}t}H'_{I}e^{-iH'_{0}t}
     =\Omega[e^{-i(\delta-\Delta)t}|0\rangle\langle1|X(t)+e^{i(\delta-\Delta)t}|1\rangle\langle0|X(t)^{+}]\\
    &+T_{c}[e^{-i\omega_{12}t}|2\rangle\langle1|X^{2}(t)+e^{i\omega_{12}t}|1\rangle\langle2|X^{2}(t)^{+}],
\end{split}
\end{equation}
where $\omega_{ij}=\varepsilon_{i}-\varepsilon_{j}$,
$\delta=\omega_{10}-\omega_{c}$ is the detuning of the applied
field to the the left dot, and
\begin{equation}
X(t)=exp[-\sum_{k}\frac{g_{k}}{2\omega_{k}}(b_{k}^{+}e^{i\omega_{k}t}-b_{k}e^{-i\omega_{k}t})],
\end{equation}
\begin{equation}
X(t)^{+}=X(t).
\end{equation}

In what follows, we assume that relaxing time of the environment
(phonon fields) is so short that the excitons do not have time to
exchange the energy and information with the environment before the
environment returns to its equilibrium state. The excitons interact
weakly with the environment so that the equilibrium thermal
properties of the environment are preserved. Therefore it is
reasonable to replace the operators $X(t)$, $X^{2}(t)$, $X(t)^{+}$
and $X^{2}(t)^{+}$ with their expectation values over the phonon
number states which are determined by a thermal average and write
the Hamiltonian as
\begin{equation}
\begin{split}
H_{eff}=&\Omega[e^{-i(\delta-\Delta)t}|0\rangle\langle1|+e^{i(\delta-\Delta)t}|1\rangle\langle0|]e^{-(N_{ph}+\frac{1}{2})\lambda}\\
        &+T_{c}[e^{-i\omega_{12}t}|2\rangle\langle1|+e^{i\omega_{12}t}|1\rangle\langle2|]e^{-2(N_{ph}+\frac{1}{2})\lambda},
\end{split}
\end{equation}
where $\lambda=\sum_{k}(g_{k}/2\omega_{k})^2$ is the Huang-Rhys
factor which corresponds to the electron-phonon
interactions\cite{huang}. As a result of quantum lattice
fluctuations, the exciton-phonon interaction affects our quantum
system even at zero temperature. Here we have made an assumption
that all the phonons have the same frequency, i.e.,
$\omega_{k}=\omega_{0}$, and write the phonon populations as
$N_{ph}=\frac{1}{e^{\omega_{0}/T}-1}$\cite{book}\cite{zhuuu}\cite{temp}.

Now we proceed to solve the equation of motion for the single
electron state vector$|\phi(t)\rangle$, i.e.,
\begin{equation}
i\frac{d}{dt}|\phi(t)\rangle=H_{eff}|\phi(t)\rangle.
\end{equation}
In general, $|\phi(t)\rangle$ is a linear combination of
$|0\rangle$, $|1\rangle$ and $|2\rangle$,
\begin{equation}
|\phi(t)\rangle=c_{0}(t)|0\rangle+c_{1}(t)|1\rangle+c_{2}(t)|2\rangle.
\end{equation}

\begin{figure}
\begin{center}
\includegraphics[scale=0.6]{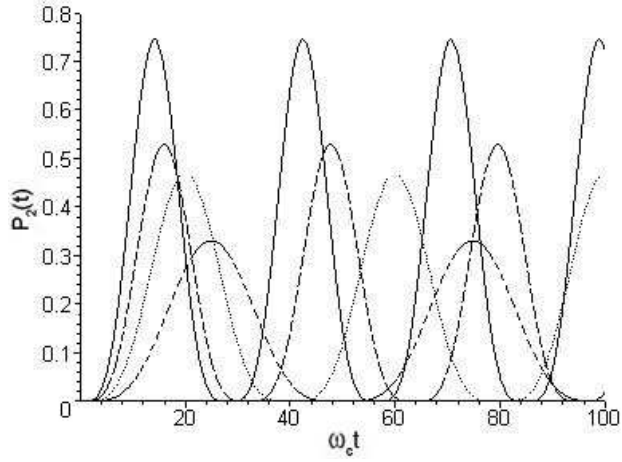}
\caption{The time evolution of the population in state $|2\rangle$
in the case $\delta=\Delta$ and $\omega_{12}=0$, for the parameters
$\Omega=0.2\omega_{c}$, $T_{c}=0.1\omega_{c}$ and $\lambda=0.01$.
Solid line is the result for $T=0$, dash line for $T=10\omega_{0}$,
dot line for $T=30\omega_{0}$ and dash dot line for
$T=50\omega_{0}$.}\label{figure2}
\end{center}
\end{figure}

\begin{figure}
\begin{center}
\includegraphics[scale=0.6]{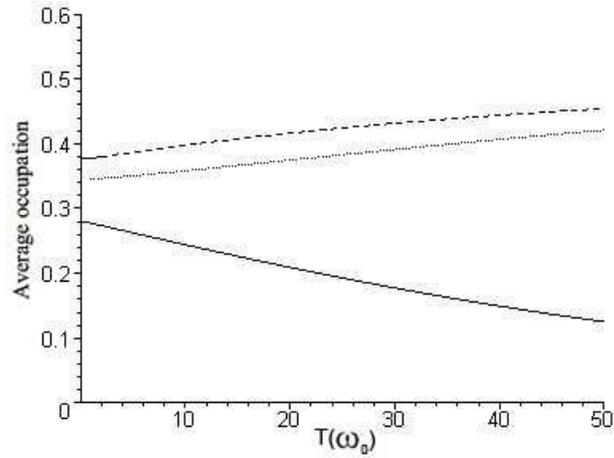}
\caption{Average occupation of state $|0\rangle$ (solid line), state
$|1\rangle$ (dash dot line) and state $|2\rangle$ (dot line) as a
function of temperature in the case $\delta=\Delta$ and
$\omega_{12}=0$, for the parameters $\Omega=0.2\omega_{c}$,
$T_{c}=0.1\omega_{c}$ and $\lambda=0.01$.}\label{figure3}
\end{center}
\end{figure}

Suppose the initial state of the system is in its ground state
$|0\rangle$, i.e., $|0\rangle=1$ and $|1\rangle=|2\rangle=0$. We
have
\begin{subequations}
\begin{equation}
i\frac{d}{dt}c_{0}(t)=\Omega{}e^{-(N_{ph}+\frac{1}{2})\lambda}c_{1}(t)e^{-i(\delta-\Delta)t},
\end{equation}
\begin{equation}
i\frac{d}{dt}c_{1}(t)=\Omega{}e^{-(N_{ph}+\frac{1}{2})\lambda}c_{1}(t)e^{i(\delta-\Delta)t}
+T_{c}e^{-2(N_{ph}+\frac{1}{2})\lambda}c_{2}(t)e^{i\omega_{12}t},
\end{equation}
\begin{equation}
i\frac{d}{dt}c_{2}(t)=T_{c}e^{-2(N_{ph}+\frac{1}{2})\lambda}c_{2}(t)e^{-i\omega_{12}t}.
\end{equation}
\end{subequations}
Assuming the optical pulse has a broad square shape, the case
$\delta=\Delta$ (the applied optical field is in resonance with the
zero-phonon line) and $\omega_{12}=0$ (state $|1\rangle$ and
$|2\rangle$ are in exactly resonance) yields.
\begin{subequations}
\begin{equation}
P_{0}(t)=|sin^{2}\theta'cos(\Theta't)+cos^{2}\theta'|^{2},
\end{equation}
\begin{equation}
P_{1}(t)=|sin\theta'sin(\Theta't)|^{2},
\end{equation}
\begin{equation}
P_{2}(t)=sin^{2}\theta'cos^{2}\theta'|cos(\Theta't)-1|^{2}.
\end{equation}
\end{subequations}
where $P_{i}=|c_{i}(t)|^{2}$ stands for the possibility of state
$|i\rangle$, $\Theta'=\sqrt{\Omega'{}^{2}+T'{}_{c}^{2}}$, and
$cos\theta'=T'_{c}/\Theta'$, with
$\Omega'=\Omega{}e^{-(N_{ph}+\frac{1}{2})\lambda}$ and
$T'_{c}=T_{c}e^{-2(N_{ph}+\frac{1}{2})\lambda}$. This analytical
result indicates that, in this simplest case, the coherent
population oscillation are similar to those at zero temperature. But
the Rabi oscillation frequency $\Omega$ and the electron-tunneling
matrix element $T_{c}$ are renormalized by the factor of
$exp[-(N_{ph}+\frac{1}{2})\lambda]$ and
$exp[-2(N_{ph}+\frac{1}{2})\lambda]$ in stead of
$exp(-\frac{1}{2}\lambda)$ and $exp(-\lambda)$, respectively, as
shown in Fig.\ref{figure2}. We note that with the increase of
temperature, the Rabi frequency drops with the decline of the
amplitude. Hence we plot in Fig.\ref{figure3} the average occupation
of state $|i\rangle$ $[1/t_{\infty}\int_{0}^{t_{\infty}}P_{i}(t)dt]$
$(i=0,1,2)$ as function of the temperature, which is obviously show
that it is less possible to create the state with one electron in
the second dot (state $|2\rangle$) when the temperature is high.

On the other hand, the environment temperature will also affect the
beat pattern, which is caused by the electron-phonon coupling. with
the rise of the temperature, the beat pattern will decay and the
population of the state $|2\rangle$ will decrease as well. As the
temperature reachs about $50\omega_{0}$, the beat pattern almost
disappears. So the environment temperature should be considered in
the practical applications.

In conclusion, we have investigated the effect of the environment
temperature on an asymmetrical double quantum dot driven by an
optical pulse. The Rabi oscillation frequency $\Omega$ and the
electron-tunneling matrix element $T_{c}$ are renormalized by the
factor of $exp[-(N_{ph}+\frac{1}{2})\lambda]$ and
$exp[-2(N_{ph}+\frac{1}{2})\lambda]$. With the rise of the
environment temperature, the population of state $|2\rangle$
declines dramatically and the beat pattern, which is caused by the
phonon effect, decays.
\begin{figure}
\begin{center}
\includegraphics[scale=0.4]{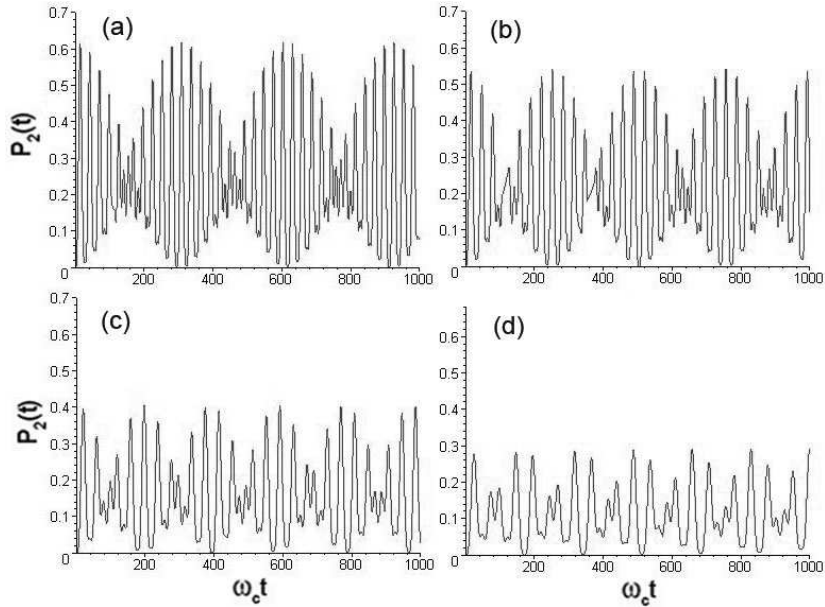}
\caption{The time evolution of the population in state $|2\rangle$,
for the parameters $\delta=\omega_{12}=0$, $\Omega=0.2\omega_{c}$,
$T_{c}=0.1\omega_{c}$, $\lambda=0.01$ and $\Delta=0.05\omega_{c}$.
(a) is the result for $T=0$, (b) for $T=10\omega_{0}$, (c) for
$T=30\omega_{0}$ and (d) for $T=50\omega_{0}$}\label{figure4}
\end{center}
\end{figure}

\end{document}